\documentclass[final,3p,times,twocolumn]{elsarticle}
\usepackage{amssymb}
\usepackage{amsfonts}
\usepackage{amsmath}
\usepackage{graphics}
\usepackage{graphicx}
\usepackage{subfigure}
\usepackage{dcolumn}
\usepackage{bm}
\usepackage{color}
\usepackage{epstopdf}
\usepackage[nodots]{numcompress}

\begin{document}

\begin{frontmatter}



\title{Zeptosecond Electron Pulse Train and Ultrafast Coherent Control of
Quantum States via Multiphoton Inelastic Cherenkov Diffraction}


\author[label1]{H. K. Avetissian\corref{cor1}}
\ead{avetissian@ysu.am}
\author[label1]{G. F. Mkrtchian}
\cortext[cor1]{Corresponding author}

\affiliation[label1]{organization={Centre of Strong Fields Physics at Research Institute of Physics, Yerevan State University},
            city={Yerevan},
            postcode={0025}, 
            country={Armenia}}

\begin{abstract}
We investigate the quantum dynamics of fermionic particles interacting with
a laser field in a gaseous medium, in the regime of inelastic diffraction
scattering on the phase lattice of a slowed travelling wave, below the
critical field of induced Cherenkov process. Using a relativistic quantum
kinetic framework and numerical solutions of Dirac equation in the rest
frame of the slowed wave, we analyze the evolution of actual electron wave
packets and beams at the inelastic scattering on the actual laser pulses of
finite duration. Our results reveal coherent multiphoton exchange involving
up to $10^{4}$ photons and the emergence of attosecond-zeptosecond electron
sub-bunches after the free-space propagation. The pulse compression by such
mechanism is robust to laser pulse duration but sensitive to the initial
momentum spread of the particles/beams. We propose a mechanism to achieve
electron pulses in zeptosecond time scales with potentiality for ultrafast
coherent control of quantum states that opens new avenues in high-resolution
temporal structuring of electron beams for time-resolved quantum
technologies and attosecond-zeptosecond science, as well as, for application
in high-resolution electron microscopy.
\end{abstract}


\end{frontmatter}



\section{Introduction}

The problem of coherent control of quantum states in physical systems
largely predicts the rapid development of quantum technologies. The
proposals to use light fields to prepare and manipulate quantum states
--particularly of free particles and their spin degrees of freedom-- trace
back to the early days of quantum mechanics \cite%
{kapitza1933reflection,rabi1937space}. However, its further development
conditioned by the advent of lasers, developed very quickly for bound-bound
(atomic/molecular) transitions, while the free-free transitions remain
significantly more challenging due to, at first, the necessity of both
accelerator (for ultrarelativistic particle beams) and laser technics, and
second - because of small electron-photon interaction cross-section compared
to atom-photon interaction one (the latter is proportional to $r_{cl}^{2}$,
where $r_{cl}=e^{2}/mc^{2}$ is classical radius of the electron $\sim
10^{-13}cm$, while the cross-section of atom-photon interaction $\sim $ $%
a^{2}$, where $a$ $\sim 10^{-8}cm$ is atomic size). Moreover, the
conservation laws for electron-photon interaction require a third body to
enable real energy-momentum exchange between the free electrons and photons.
As a result, many quantum-mechanical effects predicted for free electrons 
\cite{avetissian2006} have yet to be fully observed. It is enough to note
that the multiphoton absorption-emission in the induced free-free
transitions for the first time experimentally observed only in the late
1970s (in stimulated bremsstrahlung) \cite{weingartshofer1977direct}.

Among the various electromagnetic (EM) radiation mechanisms by free
electrons, the Cherenkov effect \cite{cerenkov1937visible,tamm1937coherent}
occupies a unique position as Cherenkov radiation emits a uniformly moving
charge in a medium with refractive index $n(\omega )>1$, at the velocity
exceeding the phase velocity of a EM wave (v $>c/n(\omega )$). So that, the
cross-section of such radiation is independent of the particle mass in
contrast to the other type -common mechanisms of EM radiation conditioned by
acceleration of a charge (so that depending on the mass of a charge). The
cross-section of Cherenkov radiation depends only on the charged particle
velocity, and the coherent length of such radiation is, in principle,
infinite. These properties make the stimulated Cherenkov process a
fundamental and versatile platform for induced free-free transitions with
relativistic particle beams and has been explored extensively as effective
mechanism for the free electron lasers (FELs)\ on ultrarelativistic
accelerator beams \cite{a1972emission,avetissian2006,freund1992principles}.
However, the most important feature of considering type radiation with
infinite coherent length concerns the stimulated Cherenkov process under
external driving EM wave. Thus, the stimulated Cherenkov effect has a strict
(both linear and nonlinear by an external wave field) peculiarity of
threshold nature connecting with the character of Cherenkov resonance in the
driving wave field. In this induced process a critical value of the
stimulating wave exists, above which the travelling wave becomes a potential
barrier or a potential well for the particle, which results to number of
field-driven resonant classical and quantum phenomena: particle "reflection"
or capture by the barrier or a well \cite%
{A1972reflection,avetissian2007nonlinear}, quantum modulation at high x-ray
frequencies \cite{avetissian1973quantum} and x-ray FEL \cite%
{avetissian2001coherent}, the formation of momentum- and energy-zone
structures in stimulated Cherenkov process \cite%
{avetissian1998quantum,avetissian1998quantum2}, etc. Below the critical
field, the quantum effects of particles inelastic diffraction on a slowed
travelling EM wave \cite{avetissian1976diffraction} (in contrast to elastic
diffraction effect on a standing wave in vacuum \cite{kapitza1933reflection}%
), modulation of an electron probability density \cite%
{avetissian1977cerenkov} have been revealed. For a comprehensive review of
the theory and applications of stimulated Cherenkov interaction, we refer
the reader to \cite{avetissian2015relativistic} and explored in the context
of FEL to \cite{freund1992principles}.

In the last two decades the significant advances in ultrafast lasers and
electron microscopy, especially through techniques like photon-induced
near-field electron microscopy (PINEM) \cite%
{barwick2009photon,park2010photon,garcia2010multiphoton}, have reignited
interest in application electron-laser interaction to shape electron beams
on ultrashort time scales. In particular, coherent phase modulation by
intense optical fields can induce multiphoton scattering \cite%
{feist2015quantum}, which enables the contribution of new approaches and
schemes in ultrafast electron microscopy, including imaging with attosecond
temporal resolution \cite{de2008electron,garcia2010optical,garcia2010optical}%
, the generation of attosecond electron pulse trains \cite%
{priebe2017attosecond,morimoto2018diffraction}, and proposals for
information encoding \cite{reinhardt2021free,gover2020free}. Furthermore,
phase-modulated electron wave packets have been proposed as a tool for
resonant coupling with bound electronic states \cite%
{gover2020free,zhang2021quantum,zhao2021quantum}.

In parallel with the above approaches, interest in the spontaneous Cherenkov
effect has seen renewed over the past decade \cite%
{kaminer2016quantum,ivanov2016quantum,roques2018nonperturbative,karataev2022observation,bae2022development,karlovets2025attosecond}%
, driven in part by its potential applications in attosecond science \cite%
{karlovets2025attosecond}. The significant advancement in attoscience is
largely connected with the realization of high-order harmonics generation
(HHG) phenomenon \cite%
{agostini1979free,corkum1993plasma,lewenstein1994theory} in atomic systems, in
the result of which ultrashort light pulses of attosecond duration have been
generated \cite{agostini2024nobel,krausz2024nobel,l2024nobel}.

Here a parallel can be drawn between the generation of attosecond photon
pulse trains by HHG in bound-bound atomic transitions and proposed in the
current paper ultrashort particle (matter waves) pulse trains in zeptosecond
time scales, generalizing the light and matter waves as quantum coherent
ensembles of photons and fermion particles as high density coherent
ensembles in ultrashort time scales for diverse applications in quantum
physics and technologies.

In the current work, on the base of the quantum theory we investigate the
stimulated Cherenkov interaction of charged fermionic particles with an
driving laser pulses in the inelastic diffraction regime, below the critical
field, considering electron wave packets and beams under laser pulses of
finite-duration with the emerging effects of kinetic instabilities due to
particles momentum spreads in wave packets/actual beams. Using a
relativistic quantum kinetic approach and direct numerical solutions of the
Dirac equation (in the rest frame of the wave, where the physical picture of
inelastic diffraction is very simple), we analyze the momentum distribution
and formation of phase-space-localized sub-pulses structures.

This paper is organized as follows. In Sec. II, the relativistic quantum kinetic 
ansatz is formulated and the results for monochromatic EM wave are presented. 
In Sec. III, we represent the results for finite laser pulses. Conclusions 
are given in Sec. IV. Appendix A represents the transition matrix elements and 
solution of Heisenberg equation. In Appendix B we consider a Gaussian laser beam 
and its shape in the rest frame of the slowed wave. Appendix C represents the 
conditions for smooth turn on and off the interaction.

\section{Relativistic quantum kinetic ansatz}

Let us consider the multiphoton interaction of charged fermionic particles
with a linearly polarized plane EM wave in a gaseous medium (see Fig. 1). The EM wave is
characterized by a carrier frequency $\omega $ and a vector potential given
by $\mathbf{A}=\boldsymbol{\epsilon }A_{e}(t,\mathbf{r})\sin \left( \omega t-%
\mathbf{kr}\right) $, where $A_{e}(t,\mathbf{r})$ is a slowly varying
amplitude, $\mathbf{k}$ is the wave vector and $\boldsymbol{\epsilon }%
=\left( 0,0,1\right) $ -unit polarization vector, $\boldsymbol{\epsilon k}=0$%
. The wave satisfies the dispersion condition $\omega ^{2}-c^{2}\mathbf{k}%
^{2}=\omega ^{2}\left( 1-n_{0}^{2}\right) <0$, where $c$ is the light speed
in vacuum, $n_{0}\equiv n(\omega )$ is the refractive index of the medium at
the carrier frequency. The wave propagation direction is given by the unit
vector $\boldsymbol{\nu }_{0}=(1,0,0)$ and $\mathbf{k}=\boldsymbol{\nu }%
_{0}n_{0}\omega /c$.

To investigate the particles/beam dynamics, we employ the quantum kinetic
approach using the second-quantized formalism of QED for considering
fermionic particles (electrons-positrons) field, where we neglect small
antiparticle (positrons) contributions on the initially given electrons
field. Furthermore, we restrict the EM wave field strength. This is a
crucial factor in the stimulated Cherenkov process due to the existence of a
critical field ($A_{cr}(t,\mathbf{r})$), above which no matter how the field
($A_{\max }(t,\mathbf{r})$) weak is, a EM wave becomes a potential barrier
for a particle \cite{A1972reflection,avetissian2007nonlinear}, as mentioned
above, and the considering diffraction regime will not take place. To this
end, we will introduce a dimensionless relativistic invariant parameter of
the wave field $\xi =eA/mc^{2}=\mathrm{inv}$ ($e$ is the charge and $m$ is
the mass of a fermionic particle) for which we suppose $\xi _{\max }<\xi
_{cr}\ll 1$ (to exclude the ionization of the dielectric medium too). 
\begin{figure}[tbp]
\includegraphics[width=0.43\textwidth]{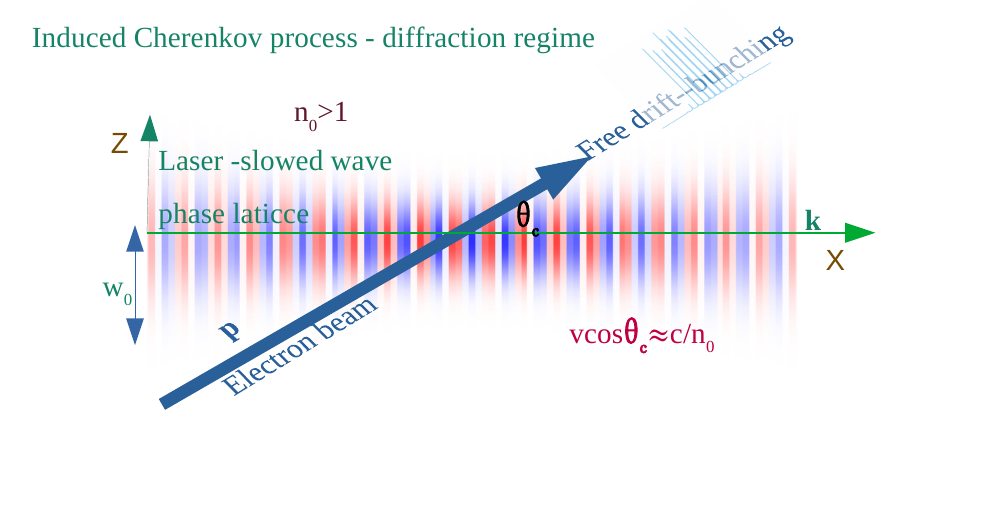}
\caption{Schematic setup. An electron with momentum $\mathbf{p}$ is incident
at an angle $\protect\theta _{c}$ onto a slowed traveling wave in a gaseous (%
$n_{0}>1$) medium. When the Cherenkov resonance condition $\mathrm{v}%
_{0}\cos \protect\theta _{c}=c/n_{0}$ is fulfilled, the traveling wave
appears as a diffraction lattice. The coherent interaction time is set by
the transverse laser width. The electron, initially modeled as a Gaussian
wave packet, undergoes multiphoton absorption/emission of laser photons and
transforms into a train of narrow pulses spaced by the phase lattice period.}
\end{figure}
The second quantized interaction Hamiltonian can be expressed in the form 
\begin{equation}
\widehat{H}_{int}=\sum_{\mathbf{p},\sigma ,\sigma ^{\prime }}\frac{ieA_{e}}{%
2c}M_{\mathbf{p,}\sigma ;\mathbf{p}-\hbar \mathbf{k,}\sigma ^{\prime
}}e^{-i\Delta \left( \mathbf{p}\right) t}\widehat{a}_{\mathbf{p},\sigma
}^{\dagger }\widehat{a}_{\mathbf{p}-\hbar \mathbf{k},\sigma ^{\prime }}+%
\mathrm{h.c.}.  \label{10}
\end{equation}%
Here the creation and annihilation operators, $\widehat{a}_{\mathbf{p}%
,\sigma }^{+}$ and $\widehat{a}_{\mathbf{p},\sigma }$, associated with
positive energy $\mathcal{E}=\sqrt{c^{2}\mathbf{p}^{2}+m^{2}c^{4}}$
solutions, satisfy the anticommutation rules at equal times, $M_{\mathbf{p}%
^{\prime }\mathbf{,}\sigma ^{\prime };\mathbf{p,}\sigma }$ is the transition
matrix element given in Appendix A, and $\Delta \left( \mathbf{p}\right) $
is Cherenkov quantum resonance detuning including quantum recoil. For
spin-preserving transitions, the matrix element is given by $M_{\mathbf{p,}%
\sigma ;\mathbf{p}\pm \hbar \mathbf{k,}\sigma }\simeq c^{2}\mathbf{\epsilon
p/}\mathcal{E=}\mathbf{\epsilon v}\delta _{\sigma \sigma ^{\prime }}$ (see
Appendix A), while for spin-flip transitions we have: $\left\vert M_{\mathbf{%
p,}\sigma ;\mathbf{p}\pm \hbar \mathbf{k,-}\sigma }\right\vert \simeq
mc^{3}\hbar \omega \mathbf{/}2\mathcal{E}^{2}$. Since due to the medium
dispersion law Cherenkov radiation takes place for optical photons $\hbar
\omega <<\mathcal{E}$, and at the condition $\left\vert \mathbf{\epsilon p}%
\right\vert \mathbf{/}mc>>\hbar \omega \mathbf{/}\mathcal{E}$ the spin-flip
transitions can be safely neglected. Accordingly, the corresponding
interaction Hamiltonian becomes%
\begin{equation}
\widehat{H}_{int}=\sum_{\mathbf{p}\sigma }\frac{ieA_{e}\mathbf{\epsilon v}}{%
2c}e^{-i\Delta t}\widehat{a}_{\mathbf{p,}\sigma }^{\dagger }\widehat{a}_{%
\mathbf{p}-\hbar \mathbf{k,}\sigma }+\mathrm{h.c.},  \label{10f}
\end{equation}%
where $\Delta =\omega -k\mathrm{v}_{x}$ is Cherenkov classical resonance
detuning. We will use Heisenberg representation, where evolution of the
operators are given by the equation 
\begin{equation}
i\hbar \frac{\partial \widehat{L}}{\partial t}=\left[ \widehat{L},\widehat{H}%
_{int}\right] ,  \label{Heis}
\end{equation}%
and expectation values are determined by the initial density matrix $%
\widehat{D}$, that is: $<\widehat{L}>=Sp\left( \widehat{D}\widehat{L}\right) 
$. From the Heisenberg equation (\ref{Heis}) with Hamiltonian (\ref{10f}),
we derive the equation of motion for the annihilation operator: 
\begin{equation}
\frac{\partial \widehat{a}_{\mathbf{p,}\sigma }}{\partial t}=\frac{eA_{e}%
\mathbf{\epsilon v}}{2\hbar c}(e^{-i\Delta t}\widehat{a}_{\mathbf{p}-\hbar 
\mathbf{k,}\sigma }-e^{i\Delta t}\widehat{a}_{\mathbf{p}+\hbar \mathbf{k,}%
\sigma }).  \label{main}
\end{equation}%
After detailed calculations (see Appendix A) involving recurrence relations
for Bessel functions $J_{n}$ and their derivatives, the exact solution of
Eq. (\ref{main}) is obtained as: 
\begin{equation}
\widehat{a}_{\mathbf{p,}\sigma }=\sum\limits_{n}\widehat{a}_{\mathbf{p+}%
n\hbar \mathbf{k,}\sigma }(0)J_{n}\left[ Z_{B}\right] e^{in\frac{\Delta }{2}%
t_{f}},  \label{sol}
\end{equation}%
where $t_{f}$ is the interaction time. The argument of Bessel function for
arbitrary detuning and wave constant amplitude is 
\begin{equation}
Z_{B}=\frac{2e\mathbf{\epsilon v}A_{e}}{\hbar c\Delta }\sin \frac{\Delta }{2}%
t_{f}.  \label{Zb}
\end{equation}%
For an arbitrary amplitude $A_{e}$, see Appendix A for details. For a single
particle initially described by a de Broglie wave with momentum $\mathbf{p}%
_{0}$ at the exact resonance ($\Delta =0$) and polarization $\sigma _{0}$
from Eq. (\ref{sol}) one can obtain the final state amplitude with momentum $%
\mathbf{p}_{0}-n\hbar \mathbf{k}$, to be $C_{\mathbf{p}_{0}-n\hbar \mathbf{k,%
}\sigma _{0}}=J_{n}\left[ Z_{B}\right] $. The latter coincides with the
result obtained in Ref. \cite{avetissian1976diffraction} (note in this
context that after several decades from the discovery of this phenomenon, a
group of authors have repeated the Cherenkov diffraction effect, making
gross errors, up to consideration of Cherenkov effect in plasma,
mixing/confusing the phase and group velocities of the wave etc., about
which see the following paper \cite{avetissian2016electron}). The argument
of the Bessel function at the exact resonance that governs the degree of
multiphotonity, can be represented in the form $Z_{B}=eE_{0}d_{\bot }/\hbar
\omega $, where $d_{\bot }=\mathrm{v}t_{f}\sin \theta _{c}$ is the coherent
interaction length. That is, $Z_{B}$ is the work done by the wave electric
field on the coherent interaction length, in units of photon energy. Note
that $Z_{B}$ does not depend on particle mass, in accordance with the stated
in the introduction fact connected with the specific case of Cherenkov
radiation at the uniform motion of a charge in a dielectric medium.

Using Eq. (\ref{sol}), one can compute the momentum-space density matrix
after the interaction ($t>t_{f}$): $\rho _{\sigma \sigma }\left( \mathbf{p}%
^{\prime },\mathbf{p,}t\right) =<\widehat{a}_{\mathbf{p}^{\prime }\mathbf{,}%
\sigma }^{+}\widehat{a}_{\mathbf{p,}\sigma }e^{i/\hbar \left( \mathcal{E}%
\left( \mathbf{p}^{\prime }\right) -\mathcal{E}\left( \mathbf{p}\right)
\right) t}>$ . Summing over the spin indices, the total density matrix will
be given by expression 
\begin{equation}
\begin{aligned} &\rho \left( \mathbf{p}^{\prime },\mathbf{p,}t\right)
=\sum\limits_{n}\sum\limits_{n^{\prime }}e^{in\Delta \left(
\mathbf{p}^{\prime }\right) t_{f}}e^{-in^{\prime }\Delta \left(
\mathbf{p}\right) t_{f}}e^{\frac{i}{\hbar }\left( \mathcal{E}\left(
\mathbf{p}^{\prime }\right) -\mathcal{E}\left( \mathbf{p}\right) \right)
t}\\ &\times \rho \left( \mathbf{p}^{\prime }\mathbf{+}n^{\prime }\hbar
\mathbf{k},\mathbf{p+}n\hbar \mathbf{k,}0\right) J_{n^{\prime }}\left[
Z_{\mathbf{p}^{\prime }}\right] J_{n}\left[ Z_{\mathbf{p}}\right].
\end{aligned}  \label{dm}
\end{equation}%
The momentum distribution is defined as $N\left( \mathbf{p,}t\right) =\rho
\left( \mathbf{p},\mathbf{p,}t\right) $. For the beam density $n\left( 
\mathbf{r}\right) $ $=\left\langle \widehat{\Psi }^{+}(\mathbf{r}^{\prime
},t)\widehat{\Psi }(\mathbf{r},t)\right\rangle _{\mathbf{r=r}^{\prime }}$ we
will have%
\begin{equation}
\begin{aligned} &n\left( \mathbf{r}\right)
=\sum_{\mathbf{p}}\sum\limits_{n,n^{\prime }}f\left( \mathbf{p}\right)
J_{n^{\prime }}\left[ Z_{\mathbf{p}}\right] J_{n}\left[
Z_{\mathbf{p}}\right] e^{i\left( n-n^{\prime }\right) \frac{\Delta
}{2}t_{f}}\\ & \times e^{\frac{i}{\hbar }\left( \mathcal{E}\left(
\mathbf{p-}n^{\prime }\hbar \mathbf{k}\right) -\mathcal{E}\left(
\mathbf{p}-n\hbar \mathbf{k}\right) \right) t}e^{i\left( n^{\prime
}\mathbf{-}n\right) \mathbf{kr}}. \end{aligned}  \label{dr}
\end{equation}%
At the condition $\left\vert \Delta \right\vert t_{f}<<1$ \bigskip we obtain 
\begin{equation}
\begin{aligned} &n\left( \mathbf{r+v}t\right)
=\sum_{\mathbf{p}}\sum\limits_{n,n^{\prime }}f\left( \mathbf{p}\right)
J_{n^{\prime }}\left[ Z_{\mathbf{p}}\right] J_{n}\left[
Z_{\mathbf{p}}\right]\\ &\times e^{i\frac{\left( n_{0}^{2}-1\right) \hbar
\omega ^{2}}{2\mathcal{E}\left( \mathbf{p}\right) }\left( n^{\prime
2}-n^{2}\right) t}e^{i\left( n^{\prime }\mathbf{-}n\right) \mathbf{kr}}.
\end{aligned}  \label{111}
\end{equation}%
Furthermore, using Eq. (\ref{dm}), one can construct the corresponding
Wigner quasiprobability distribution $W(\mathbf{r,p},t)$ for analysis of the
coherence and localization in both position and momentum. Due to symmetry
with direction $\mathbf{\epsilon }$ ($OZ$ axis), for numerical calculations
we have taken, without loss of generality, the vector $\mathbf{p}$ in the $%
XZ $ plane $(p_{y}=0)$. Then, since $p_{\bot }=const$, we only consider
Gaussian single particle wave packets with momentum uncertainty $\delta
p_{_{x}}$, or beam with longitudinal momentum width $\Delta p_{_{x}}$.

In Fig. 2(a), we illustrate a typical signature of multiphoton
absorption-emission in the induced Cherenkov process. The final
momentum-space distribution of a single electron with initial Gaussian
packet $\delta p_{_{x}}<<\hbar k$, shows a symmetric distribution over the
photon number. Peaks emerge near $s\simeq \pm Z_{B}$, consistent with the
behavior of Bessel functions: for $Z_{B}>>1$, the latter reaches its maximum
at $\left\vert s\right\vert \simeq Z_{B}$, which dominates the contribution
to the momentum distribution. 
\begin{figure}[tbp]
\includegraphics[width=0.42\textwidth]{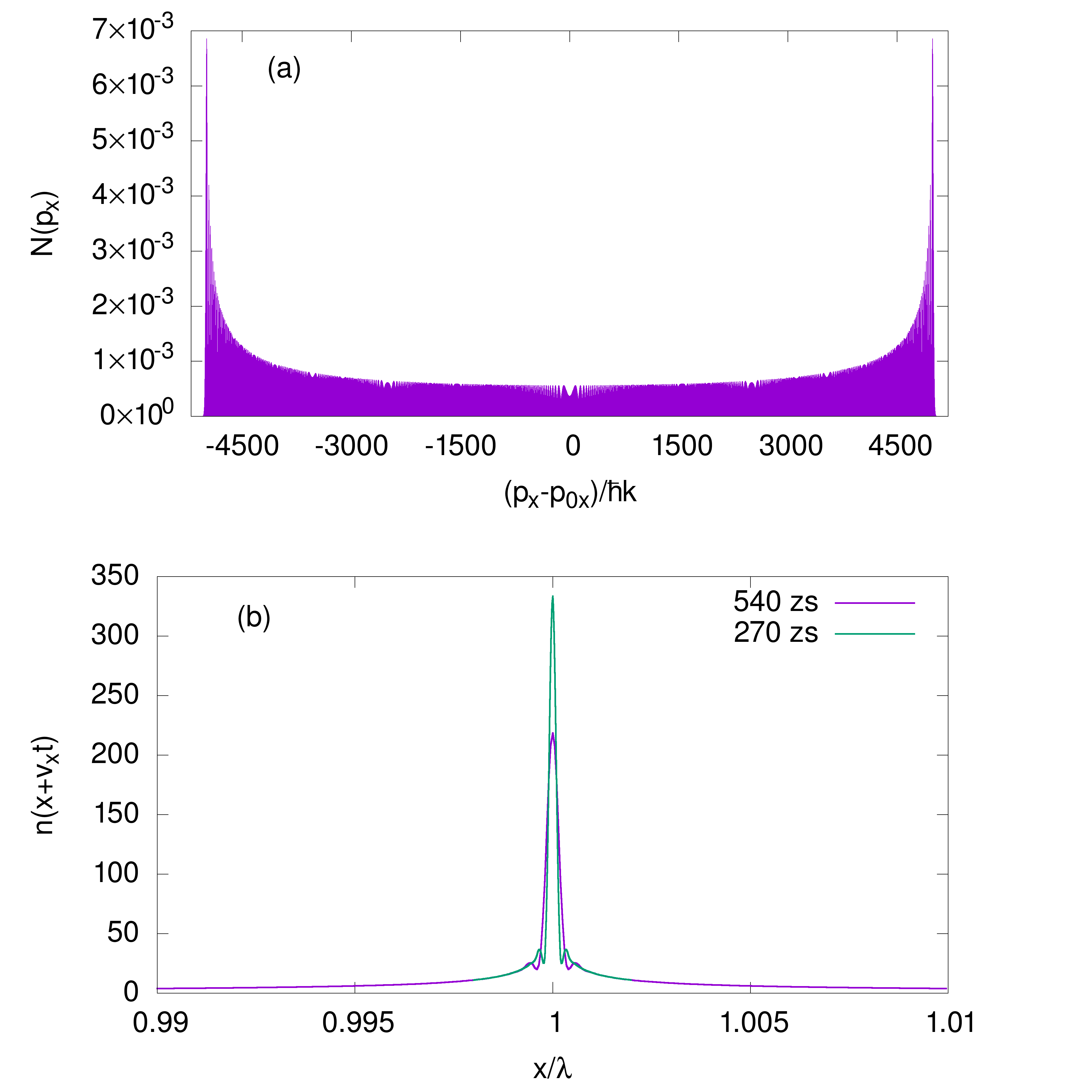}
\caption{(a) Final momentum-space distribution of a de Broglie wave with
Lorentz factor $\protect\gamma =25$ and Cherenkov angle $\protect\theta %
_{c}=1/(10\protect\gamma )$. Interaction length: $d_{\bot }=1.\,55\times
10^{-2}\,\mathrm{cm}$; laser wavelength $\protect\lambda =800\,\mathrm{nm}$;
wave electric field amplitude $E_{0}=5\times 10^{5}\,\mathrm{V/cm}$ ($%
\protect\xi _{0}=1.25\times 10^{-5}$.). (b) Formation of a zeptosecond
electron pulse train after the interaction. Free propagation time $%
t_{p}=t_{c}$. Shown is one of the emerging peaks, separated by the laser
wavelength. For electric field strengths $E_{0}=10^{6}$ $\mathrm{V/cm}$ and $%
E_{0}=5\times 10^{5}$ $\mathrm{V/cm}$, the resulting pulse durations are $%
270\ \mathrm{zs}$ and $540\ \mathrm{zs}$, respectively.}
\end{figure}
After the interaction, the momentum distribution remains unchanged, whereas
the spatial distribution evolves as described by Eq. (\ref{dr}), forming a
train of narrow peaks separated by the phase lattice period. The exponential
factor $\exp \left[ i\left( n_{0}^{2}-1\right) \hbar \omega ^{2}/2\mathcal{E}%
\left( n^{\prime 2}-n^{2}\right) t\right] $ in Eq. (\ref{111}) leads to
strong spatial bunching around the modes with $\left\vert n^{\prime }\mathbf{%
-}n\right\vert \sim Z$ after the free-space propagation. The maximum
bunching takes place at times $t\simeq t_{c}$, where%
\begin{equation}
t_{c}=\frac{1}{Z_{B}}\frac{\mathcal{E}}{\left( n_{0}^{2}-1\right) \hbar
\omega ^{2}}  \label{Tq}
\end{equation}%
is the time at which the constructive interference between the electron
states corresponding to $n$ photon absorption-emission in (\ref{111}) takes
place, which after the propagation in the free-space leads to compressed
pulse train-structure of ultrashort (zeptosecond) duration. Figure 2(b)
shows one of the emerging peaks in the electron density profile. These peaks
are spaced by the laser wavelength and correspond to pulse durations in the
zeptosecond regime. It is important to note that this is a single-particle
quantum interference effect. In actual beams, the divergence in the
longitudinal momentum (phase mismatch) leads to broadening of these peaks.
For sufficiently large values of $\Delta p_{x}$, the peaks may eventually
disappear. Therefore, understanding the extent to which these ultrashort
features survive is important. Figures 3 and 4 present numerical results for
electron beams with relative longitudinal momentum spreads of $\Delta
p_{x}/p_{0x}=10^{-6}$ and $\Delta p_{x}/p_{0x}=10^{-5}$, respectively. The
bunching effect is clearly illustrated through the Wigner quasiprobability
distribution, evaluated at time $t=t_{c}$ near one of the dominant peaks of
the propagated electron state. Initially, interaction with the laser field
imprints a periodic structure in momentum space, corresponding to a
superposition of discrete momentum components spaced by the photon momentum.
As the wavefunction propagates freely, this modulation evolves into spatial
localization due to quantum dispersion. The resulting Wigner distribution
which encodes both momentum and position information, develops sharply
localized features in phase space signaling the formation of well-defined,
ultrashort electron pulses in real space. These sub-cycle structures, with
durations in the atto- to zeptosecond range, emerge as a direct consequence
of quantum interference. As expected, the peaks become broadened with
increasing beam divergence, yet even for $\Delta p_{x}/p_{0x}=10^{-5}$,
significant temporal compression is retained, indicating the robustness of
the effect. 
\begin{figure}[tbp]
\includegraphics[width=0.42\textwidth]{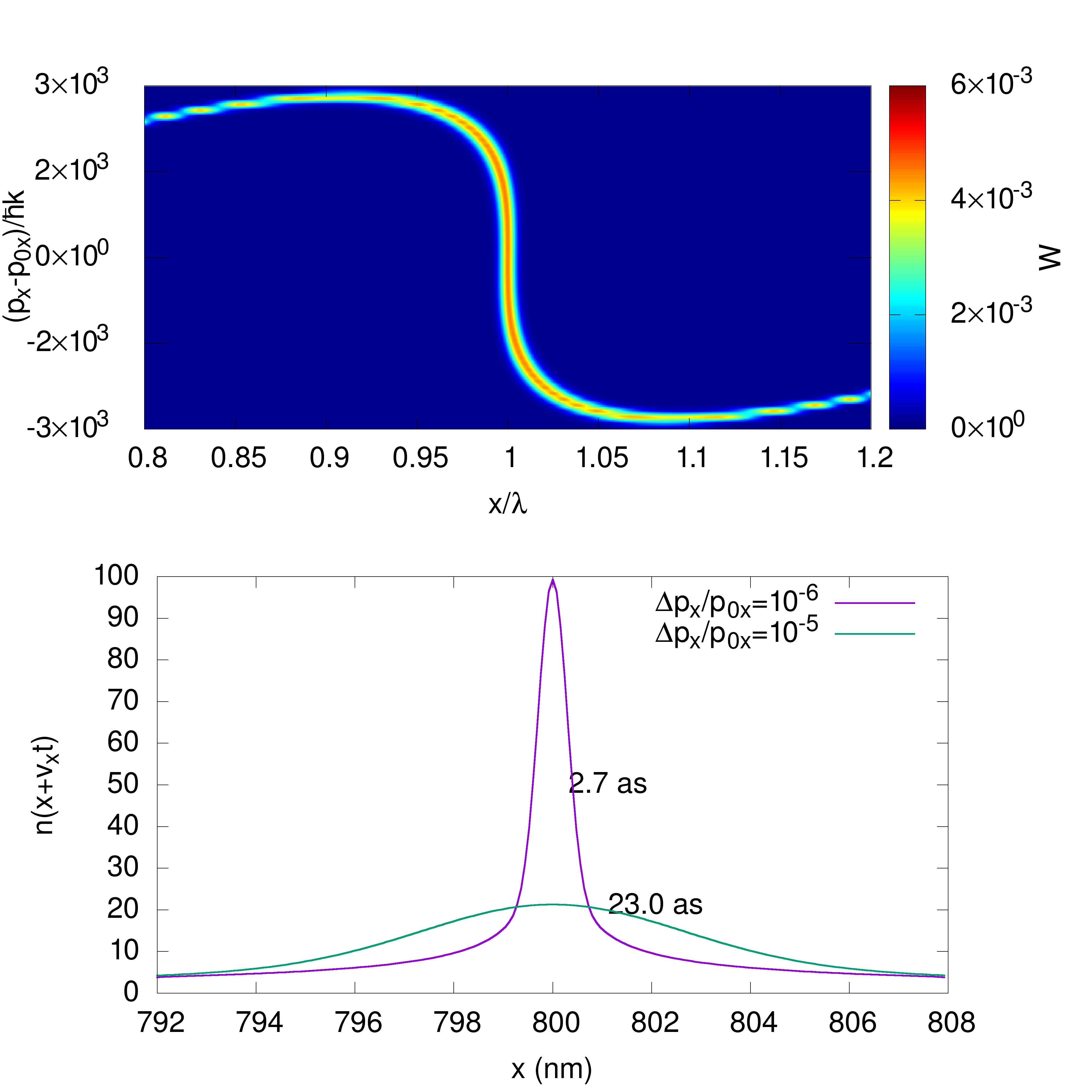}
\caption{Generation of an ultrashort electron pulse train. Top panel: Wigner
function $W(x+\mathrm{v}_{x}t\mathbf{,}p_{x}+p_{0x},t)$ at $t_{p}=t_{c}$ for
an initially uniform electron beam with a Gaussian momentum distribution of $%
\Delta p_{x}/p_{0x}=10^{-5}$. Bottom panel: Resulting electron beam density
for $\Delta p_{x}/p_{0x}=10^{-6}$ and $\Delta p_{x}/p_{0x}=10^{-5}$.
Parameters: $\protect\gamma =25$, $\protect\theta _{c}=1/(10\protect\gamma )$%
, $\protect\lambda =800\,\mathrm{nm}$, $d_{\bot }=1.\,55\times 10^{-2}%
\mathrm{cm}$, and $E_{0}=3\times 10^{5}$ $\mathrm{V/cm}$.}
\end{figure}
\begin{figure}[tbp]
\includegraphics[width=0.42\textwidth]{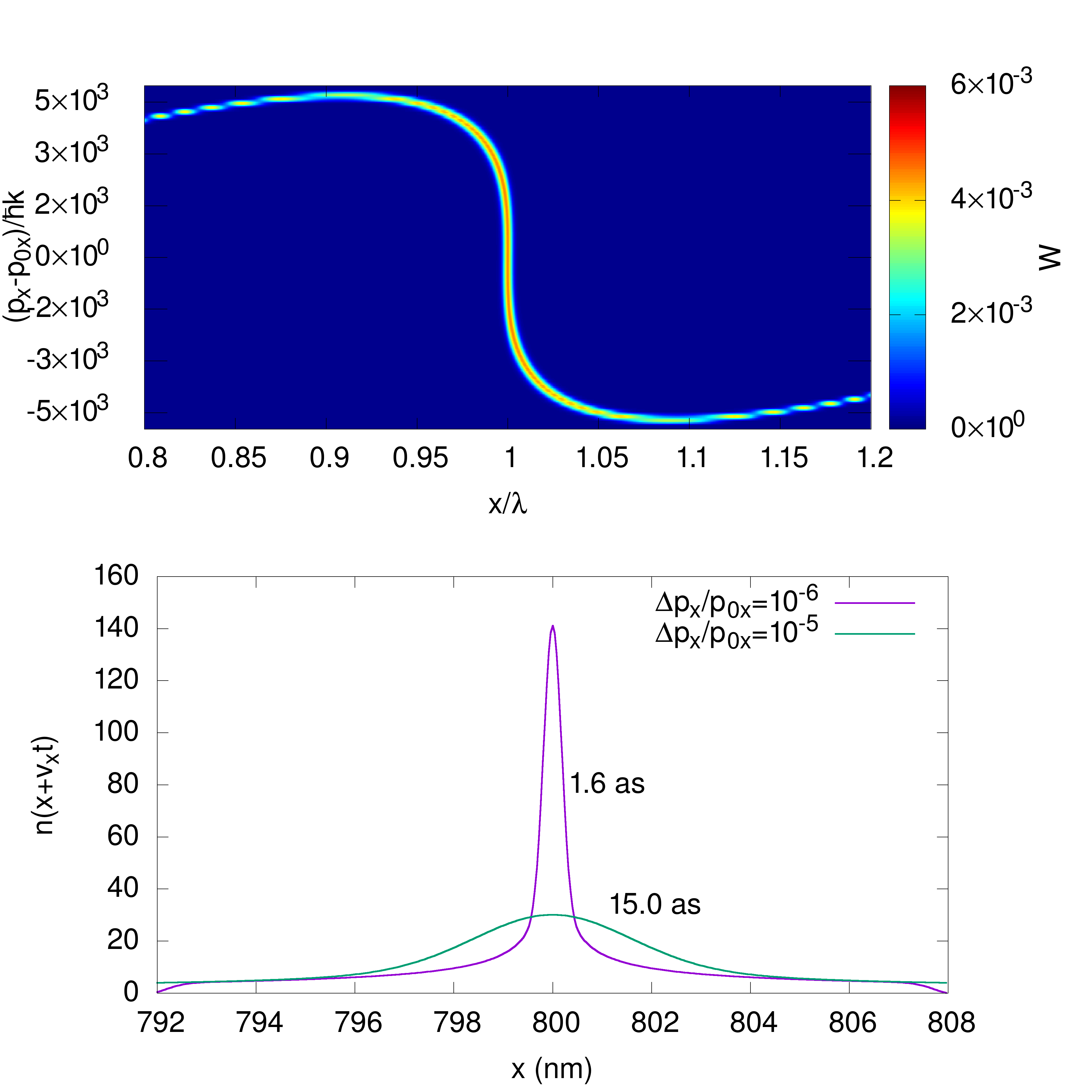}
\caption{Same as Figure 3 but for $E_{0}=5\times 10^{5}$ $\mathrm{V/cm}$.}
\end{figure}

\section{Finite pulse effects}

As discussed above, phase mismatch resulting from electron beam divergence
leads to a broadening of the emerging peaks. Additionally, these features
can be influenced by EM wave itself due to the finite duration of actual
laser pulses. To account for the impact of such finite-duration EM pulses,
we consider the dynamics of a spin-1/2 fermion governed by the Dirac
equation: 
\begin{equation}
i\hbar \frac{\partial \Psi }{\partial t}=\left[ c\widehat{\mathbf{\alpha }}(%
\widehat{\mathbf{p}}-\frac{e}{c}\mathbf{A})+\widehat{\beta }mc^{2}\right]
\Psi .  \label{f1}
\end{equation}%
To facilitate the solution of Eq. (\ref{f1}), we transform to the frame of
reference moving with the phase velocity $\mathrm{V}=c/n_{0}$ of the wave
-hereafter referred to as the wave rest ($R$ ) frame (see Appendix B). In
this frame, the radiation field appears as a quasistatic magnetic field with
the vector potential $\mathbf{A}_{R}=\left\{ 0,0,A_{0}(x,t)\sin \left(
k^{\prime }x\right) \right\} $, where $ck^{\prime }=\omega \sqrt{n_{0}^{2}-1}
$. The bispinor wave function in the $R$ frame is related to that in the
laboratory ($L$) frame via a Lorentz transformation. We solve Eq. (\ref{f1})
numerically, fully accounting for the finite pulse shape, quantum recoil
effect, and spin-flip transitions. For the wave envelope $A_{0}(x,t)$ we
assume $A_{0}(x,t)=A_{0}f(x)g\left( t\right) $. The spatial part is
described by the sin-squared function: $f(x)=\sin ^{2}\left( \pi x/\delta
\right) $, while temporal part provides smooth turn on and off the
interaction (see, Appendix C for details). Interaction parameters are chosen
to match those in the $L$ frame. Figure 5(a) shows the final momentum-space
distribution of an initially spin-up electron described by a Gaussian wave
packet with the large transverse momentum. In this regime, the dominant
interaction arises from the term $\sim \mathbf{A}_{R}\mathbf{p}_{\bot }$.
The resulting diffraction spectrum closely resembles that of interaction
with a monochromatic wave, though with slight asymmetry between the photon
absorption and emission branches. Spin-flip transitions are negligible in
this case. Figure 5(b) displays the time evolution of the electron's
probability density, illustrating the transformation of the initial Gaussian
wave packet into a sequence of sharp peaks located at the center of each
phase lattice cell, spaced by the phase lattice period. These peaks reach
their maximum at approximately $t\simeq t_{c}/\gamma $. This is further
shown in Fig. 6 for phase lattices consisting of $N_{k}=10$ and $N_{k}=20$
periods. After Lorentz transformation to the $L$ frame, the resulting
electron pulse durations correspond to approximately $1300\ \mathrm{zs}$ and 
$900\ \mathrm{zs}$, respectively. As evident from this figure, the finite
duration of the laser pulse leads to a broadening of the resulting peaks
compared to the idealized monochromatic case. Nevertheless, increasing the
number of lattice periods improves the compression effect, with optimal
results achieved around the periods $N_{k}\gtrsim 20$. 
\begin{figure}[tbp]
\includegraphics[width=0.42\textwidth]{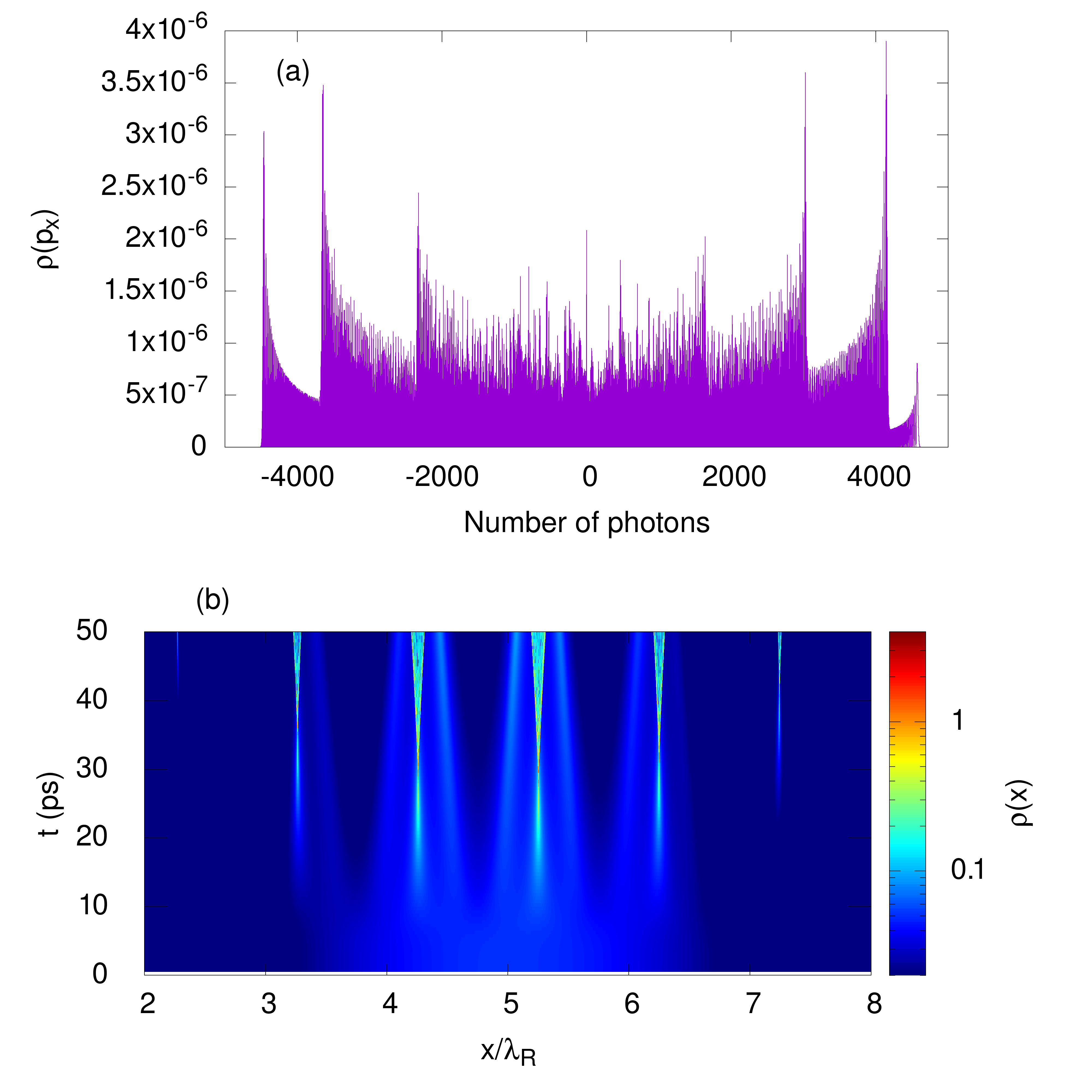}
\caption{Multiphoton absorption-emission in the $R$ frame in the induced
Cherenkov process. (a) Final momentum-space distribution of an initial
electron described by a Gaussian wave packet with position uncertainty $%
\protect\delta x=8/k^{\prime }$, momentum $p_{x}=0$, centered at $%
x_{0}=5/k^{\prime }$. The phase lattice contains $N_{k}=10$ periods.
Interaction is smoothly turned on and off using a Gaussian envelope with $%
\protect\tau _{w}=2.67\,\mathrm{ps}$. The transverse momentum is
relativistic invariant, set to $p_{z}=0.1mc$. The parameter of the
electron-wave interaction is $\protect\xi _{0}=1.25\times 10^{-5}$. (b) Time
evolution of the probability density. The color scale reveals the
transformation of the Gaussian wave packet into a sequence of narrow peaks
separated by the phase lattice period: $\protect\lambda _{R}=2\protect\pi %
/k^{\prime }$.}
\end{figure}
\begin{figure}[tbp]
\includegraphics[width=0.38\textwidth]{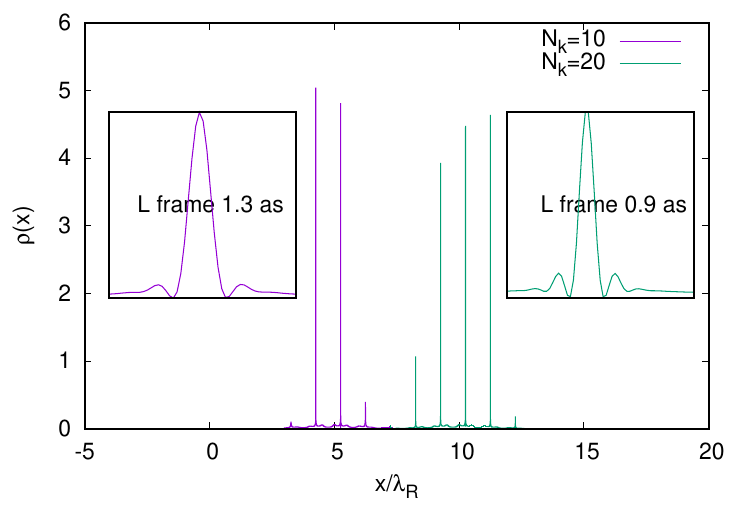}
\caption{Formation of an ultrashort electron pulse train in the $R$ frame.
Probability density at $t=28\,\mathrm{ps}$ for phase lattices with $N_{k}=10$
and $N_{k}=20$ periods. The insets show magnified views of peaks
corresponding to $1300\ \mathrm{zs}$ and $900\ \mathrm{zs}$ pulses in the
laboratory frame. Other parameters as in Fig. 5.}
\end{figure}

\section{Conclusion}

In conclusion, within the second-quantized formalism of QED for fermionic
particles (electrons-positrons) field, solving analytically the -Heisenberg
and numerically - the Dirac equations, we revealed fermion particles
(matter-wave) localization and pulse train formation in stimulated Cherenkov
process in the multiphoton inelastic diffraction regime on a slowed wave
phase lattice in a gaseous medium. Our results reveal coherent multiphoton
exchange involving up to $10^{4}$ photons and the emergence of shorter than
attosecond electron sub-bunches after the free-space propagation. The phase
modulation, initially created by a laser field, evolve through the
free-space propagation into a sequence of narrow, high-density peaks leading
to ultrashort matter waves pulse trains structure on the zeptosecond time
scale. We suggest practical routes for generation of zeptosecond electron
pulse trains, coherent control of electrons quantum states in ultrashort
time scales for applications in high-resolution electron microscopy and
time-resolved ultrafast quantum technologies. In addition, is of special
interest the creation of high density coherent ensembles of fermion
particles as "laser sources", specifically, in sub-attosecond time scales,
towards the diverse applications in quantum optics-electronics.

\section*{Acknowledgments}

The work was supported by the Science Committee of Republic of Armenia,
project No. 24WS-1C004.

\appendix

\section{Transition matrix elements and solution of Heisenberg equation}

The QED Hamiltonian is expressed as 
\begin{equation}
\widehat{H}_{int}=-\frac{1}{c}\int d\mathbf{r}\widehat{\mathbf{j}}\mathbf{A,}
\end{equation}%
with the current density operator defined as 
\begin{equation}
\widehat{\mathbf{j}}=ec\widehat{\Psi }^{+}\gamma _{0}\boldsymbol{\gamma }%
\widehat{\Psi },
\end{equation}%
where $\gamma _{0}$ and $\boldsymbol{\gamma}$ are the Dirac matrices, which
we will take in the spinor representation \cite{berestetskii2012quantum}.
The fermionic field operator $\widehat{\Psi }$ is expanded in terms of free
Dirac states $\psi _{\mathbf{p}\sigma }(\mathbf{r},t)$ as%
\begin{equation}
\widehat{\Psi }(\mathbf{r},t)=\sum_{\mathbf{p},\sigma }\widehat{a}_{\mathbf{p%
},\sigma }(t)\psi _{\mathbf{p}\sigma }(\mathbf{r},t),
\end{equation}

The free particle solutions $\psi _{\mathbf{p},\sigma }=\left( 2\mathcal{E}%
\right) ^{-1/2}u_{\sigma }\left( \mathbf{p}\right) e^{i/\hbar \left( \mathbf{%
pr-}\mathcal{E}t\right) }$ of \ Dirac equation $\left( \mathcal{E}\gamma
_{0}-c\mathbf{p\gamma }-mc^{2}\right) u_{\sigma }\left( \mathbf{p}\right) =0$
with positive energies and polarizations $\sigma =\pm \frac{1}{2}$ (spin
projections $\mathbf{\epsilon S}=\pm \frac{1}{2}$ in the rest frame of the
particle) are defined by the bispinors 
\begin{equation}
u_{1/2}\left( \mathbf{p}\right) =\sqrt{\frac{1}{\left( \mathcal{E}-c\mathbf{%
\epsilon p}\right) }}\left( 
\begin{array}{l}
mc^{2}w^{(1/2)} \\ 
\\ 
\left( \mathcal{E}-c\mathbf{\sigma p}\right) w^{(1/2)}%
\end{array}%
\right) ,  \label{B1}
\end{equation}%
\begin{equation}
u_{-1/2}\left( \mathbf{p}\right) =\sqrt{\frac{1}{\left( \mathcal{E}+c\mathbf{%
\epsilon p}\right) }}\left( 
\begin{array}{l}
\left( \mathcal{E}+c\mathbf{\sigma p}\right) w^{(-1/2)} \\ 
\\ 
mc^{2}w^{(-1/2)}%
\end{array}%
\right) ,  \label{B2}
\end{equation}%
\bigskip where $\mathcal{E}=\sqrt{c^{2}\mathbf{p}^{2}+m^{2}c^{4}}$, $\mathbf{%
\sigma }=\left( \sigma _{x},\sigma _{y},\sigma _{z}\right) $ are the Pauli
matrices and the spinors $w^{(\pm 1/2)}$ are: 
\begin{equation}
w^{(1/2)}=\left( 
\begin{array}{l}
1 \\ 
0%
\end{array}%
\right) ;\qquad w^{(-1/2)}=\left( 
\begin{array}{l}
0 \\ 
1%
\end{array}%
\right) .  \label{B4}
\end{equation}%
The transition matrix element defined in the main text are 
\begin{equation}
M_{\mathbf{p}^{\prime }\mathbf{,}\sigma ^{\prime };\mathbf{p,}\sigma }==c%
\frac{\overline{u}_{\sigma ^{\prime }}(p^{\prime })\widehat{\epsilon }%
u_{\sigma }(p)}{2\sqrt{\mathcal{E}^{\prime }\mathcal{E}}}=c\frac{u_{\sigma
^{\prime }}^{+}\left( p^{\prime }\right) \alpha _{z}u_{\sigma }(p)}{2\sqrt{%
\mathcal{E}^{\prime }\mathcal{E}}},  \label{B5}
\end{equation}%
where 
\begin{equation}
\alpha _{z}=\left[ 
\begin{array}{cccc}
1 & 0 & 0 & 0 \\ 
0 & -1 & 0 & 0 \\ 
0 & 0 & -1 & 0 \\ 
0 & 0 & 0 & 1%
\end{array}%
\right] .  \label{B6}
\end{equation}%
Taking into account Eqs. (\ref{B1}-\ref{B6}) we have 
\begin{equation*}
M_{\mathbf{p}^{\prime }\mathbf{,}1/2;\mathbf{p,}1/2}=\frac{c}{\sqrt{4%
\mathcal{E}\mathcal{E}^{\prime }\left( \mathcal{E}-cp_{z}\right) \left( 
\mathcal{E}^{\prime }-cp_{z}\right) }}
\end{equation*}%
\begin{equation}
\times \left[ \left( \mathcal{E-}cp_{z}\right) \left( \mathcal{E}-\mathcal{E}%
^{\prime }+2cp_{z}\right) +c^{2}\left( p_{x}+ip_{y}\right) \left(
p_{x}^{\prime }-p_{x}\right) \right] ,  \label{B7}
\end{equation}

\begin{equation}
M_{\mathbf{p}^{\prime }\mathbf{,}1/2;\mathbf{p,-}1/2}=\frac{mc^{4}\left(
p_{x}^{\prime }-p_{x}\right) }{\sqrt{4\mathcal{E}\mathcal{E}^{\prime }\left( 
\mathcal{E}-cp_{z}\right) \left( \mathcal{E}^{\prime }+cp_{z}\right) }}.
\label{B8}
\end{equation}%
Next we consider solution of Hesenberg equation 
\begin{equation}
i\hbar \frac{\partial \widehat{a}_{\mathbf{p}}}{\partial t}=\frac{eA_{e}%
\mathbf{\epsilon v}}{2c}e^{-i\Delta \left( \mathbf{p}\right) t}\widehat{a}_{%
\mathbf{p}-\hbar \mathbf{k}}+\frac{eA_{e}^{\ast }\mathbf{\epsilon v}}{2c}%
e^{i\Delta \left( \mathbf{p}+\hbar \mathbf{k}\right) t}\widehat{a}_{\mathbf{p%
}+\hbar \mathbf{k}},  \label{B9}
\end{equation}%
where%
\begin{equation}
\hbar \Delta \left( \mathbf{p}\right) =\sqrt{c^{2}\left( \mathbf{p}-\hbar 
\mathbf{k}\right) ^{2}+m^{2}c^{4}}-\sqrt{c^{2}\mathbf{p}^{2}+m^{2}c^{4}}%
+\hbar \omega  \label{rdet}
\end{equation}%
is the resonance detuning. Thus, from $\Delta \left( \mathbf{p}\right) =0$
we obtain%
\begin{equation*}
\left( 1-\frac{\mathbf{vk}}{\omega }\right) =-\frac{\hbar \omega }{2\mathcal{%
E}}\left( n_{0}^{2}-1\right) ,
\end{equation*}%
which is the Cherenkov resonance condition for emission of one photon taking
into account quantum recoil and $\Delta \left( \mathbf{p}+\hbar \mathbf{k}%
\right) =0$ gives resonance condition for absorption:%
\begin{equation*}
\left( 1-\frac{\mathbf{vk}}{\omega }\right) =\frac{\hbar \omega }{2\mathcal{E%
}}\left( n_{0}^{2}-1\right) .
\end{equation*}%
To avoid negative effects of multiple scattering and ionization loss of the
particle we consider the gases of relatively low densities. The optimal
values of the refractive index of the gaseous media for Cherenkov process
are $n_{0}-1\sim 10^{-3}-10^{-5}$ and frequencies $\hbar \omega =0.1-3$ eV.
Hence quantum recoil is negligibly small and can be safely neglected. In
this case resonance condition for absorption and emission are the same $1-%
\mathbf{vk/}\omega =0$. The latter is the classical resonance condition.

Introducing new operators $\widehat{a}_{\mathbf{p-}n\hbar \mathbf{k}}=%
\widehat{f}_{n}(\mathbf{p})$ and neglecting quantum recoil, for $\widehat{f}%
_{n}(\mathbf{p})$ from Eq. (\ref{B9}) we obtain%
\begin{equation*}
i\hbar \frac{\partial \widehat{f}_{n}(\mathbf{p})}{\partial t}=\frac{e%
\mathbf{\epsilon v}\left\vert A_{e}\right\vert }{2c}
\end{equation*}

\begin{equation}
\times \left[ e^{-i\left( \omega -\mathbf{vk}\right) t+i\varphi }\widehat{f}%
_{n+1}(\mathbf{p})+e^{i\left( \omega -\mathbf{vk}\right) t-i\varphi }%
\widehat{f}_{n-1}(\mathbf{p})\right] ,  \label{B10}
\end{equation}%
where $\varphi =\arg A_{e}$. Let us consider the solution of Eq. (\ref{B10})
in the form

\begin{equation}
\widehat{f}_{n}(\mathbf{p},t)=\sum\limits_{n^{\prime }}f_{n-n^{\prime }}(%
\mathbf{p},0)J_{n^{\prime }}\left[ Z\left( t\right) \right] e^{in^{\prime
}\Phi \left( t\right) }\bigskip ,  \label{B11}
\end{equation}%
where $J_{n}\left[ Z\right] $ is the Bessel function, $Z\left( t\right) $
and $\Phi \left( t\right) $ are unknown functions. Involving recurrence
relations for Bessel functions $J_{n}$ and their derivatives

\begin{eqnarray*}
\frac{d}{dZ}J_{N}\left[ Z\right] &=&\frac{1}{2}J_{N-1}\left[ Z\right] -\frac{%
1}{2}J_{N+1}\left[ Z\right] , \\
J_{N}\left[ Z\right] &=&\frac{Z}{2N}\left( J_{N+1}\left[ Z\right] +J_{N-1}%
\left[ Z\right] \right) ,
\end{eqnarray*}%
form Eqs. (\ref{B10}) and (\ref{B11}) we obtain coupled equations: 
\begin{eqnarray}
Z^{\prime }\left( t\right) &=&\frac{e\mathbf{\epsilon v}\left\vert
A_{e}\right\vert }{\hbar c}\sin \left[ \Delta t-\varphi -\Phi \left(
t\right) \right] ,  \notag \\
\Phi ^{\prime }\left( t\right) Z\left( t\right) &=&-\frac{e\mathbf{\epsilon v%
}\left\vert A_{e}\right\vert }{\hbar c}\cos \left[ \Delta t-\varphi -\Phi
\left( t\right) \right] .  \label{czfi}
\end{eqnarray}%
These equations allow analytical solutions for two physical interesting
cases. Namely at exact resonance ($\Delta =0$) and arbitrary envelope $A_{e}$
we have 
\begin{eqnarray}
\Phi &=&-\frac{\pi }{2}-\varphi ,  \notag \\
Z &=&\frac{e\mathbf{\epsilon v}}{\hbar c}\int_{0}^{t}\left\vert
A_{e}\right\vert dt.  \label{Z1}
\end{eqnarray}%
Then for arbitrary detuning and constant envelope we obtain 
\begin{eqnarray}
\Phi &=&-\frac{\pi }{2}-\varphi +\frac{\Delta }{2}t,  \notag \\
Z &=&\frac{2e\mathbf{\epsilon v}\left\vert A_{e}\right\vert }{c}\frac{\sin %
\left[ \frac{\Delta }{2}t\right] }{\Delta }.  \label{Z2}
\end{eqnarray}%
In the main text, for concreteness, the solution (\ref{Z2}) is considered
with $\varphi =-\pi /2$.

\section{Gaussian laser beam and Lorentz transformation to the frame of
reference moving with the phase velocity of the wave}

For the linearly polarized Gaussian laser beam propagating in the $+x$
direction the electric and magnetic fields are given by the following
expressions \cite{mcdonald2000axicon}:%
\begin{equation*}
E_{z}=\frac{E_{0}}{\sqrt{1+\frac{x^{2}}{x_{R}^{2}}}}e^{-\frac{r_{\bot }^{2}}{%
w^{2}\left( x\right) }}f\left( x-\frac{\omega }{k}t\right)
\end{equation*}%
\begin{equation*}
\times \cos \left( kx-\omega t+\Phi \left( x,r_{\bot }\right) -\tan
^{-1}\left( \frac{x}{x_{R}}\right) \right)
\end{equation*}%
\begin{equation*}
E_{x}=\frac{z}{x_{R}}\frac{E_{0}}{1+\frac{x^{2}}{x_{R}^{2}}}e^{-\frac{%
r_{\bot }^{2}}{w^{2}\left( x\right) }}f\left( x-\frac{\omega }{k}t\right)
\end{equation*}%
\begin{equation}
\times \sin \left( kx-\omega t+\Phi \left( x,r_{\bot }\right) -2\tan
^{-1}\left( \frac{x}{x_{R}}\right) \right)  \label{A1}
\end{equation}%
\begin{equation*}
H_{y}=-\frac{n_{0}E_{0}}{\sqrt{1+\frac{x^{2}}{x_{R}^{2}}}}e^{-\frac{r_{\bot
}^{2}}{w^{2}\left( x\right) }}f\left( x-\frac{\omega }{k}t\right)
\end{equation*}%
\begin{equation*}
\times \cos \left( kx-\omega t+\Phi \left( x,r_{\bot }\right) -\tan
^{-1}\left( \frac{x}{x_{R}}\right) \right)
\end{equation*}%
\begin{equation*}
H_{x}=\frac{y}{x_{R}}\frac{E_{0}}{1+\frac{x^{2}}{x_{R}^{2}}}e^{-\frac{%
r_{\bot }^{2}}{w^{2}\left( x\right) }}f\left( x-\frac{\omega }{k}t\right)
\end{equation*}%
\begin{equation*}
\times \sin \left( kx-\omega t+\Phi \left( x,r_{\bot }\right) -2\tan
^{-1}\left( \frac{x}{x_{R}}\right) \right)
\end{equation*}%
where $f\left( x-\frac{\omega }{k}t\right) $ is a slowly varying envelope,%
\begin{equation}
w\left( x\right) =w_{0}\sqrt{1+\frac{x^{2}}{x_{R}^{2}}}  \label{A2}
\end{equation}%
is the transverse size of the beam at position $x$, $w_{0}$ is the waist, $%
x_{R}=kw_{0}^{2}/2$ is the Rayleigh range, and 
\begin{equation*}
\Phi \left( x,r_{\bot }\right) =\frac{r_{\bot }^{2}/x_{R}^{2}}{2(1+\frac{%
x^{2}}{x_{R}^{2}})}
\end{equation*}%
is the extra phase shift due to focusing. For a large beam waists $%
w_{0}>100\lambda $ one can safely ignore longitudial components of the
fields and phase shift $\Phi \left( x,r_{\bot }\right) $.

Making Lorentz transformation to the $R$ frame moving at velocity $\mathrm{V}%
=\omega /k=c/n_{0}$%
\begin{eqnarray*}
x &=&\gamma \left( x^{\prime }+\frac{c}{n_{0}}t^{\prime }\right) ;\ r_{\bot
}=r_{\bot }^{\prime };\  \\
t &=&\gamma \left( x^{\prime }+\frac{t^{\prime }}{n_{0}c}\right) ;\ \gamma =%
\frac{n_{0}}{\sqrt{n_{0}^{2}-1}};
\end{eqnarray*}%
\begin{eqnarray}
\mathbf{E}_{\parallel }^{\prime } &=&\mathbf{E}_{\parallel };\ \mathbf{E}%
_{\perp }^{\prime }=\gamma \left( \mathbf{E}_{\perp }+\frac{1}{c}\mathbf{%
V\times H}\right) ;  \notag \\
\mathbf{H}_{\parallel }^{\prime } &=&\mathbf{H}_{\parallel };\ ;\ \mathbf{H}%
_{\perp }^{\prime }=\gamma \left( \mathbf{H}_{\perp }-\frac{1}{c}\mathbf{%
V\times E}\right) ,  \label{A33}
\end{eqnarray}%
we obtain%
\begin{equation}
E_{z}^{\prime }=0,  \label{A4}
\end{equation}%
\begin{equation*}
H_{y}^{\prime }=-\frac{E_{0}\sqrt{n_{0}^{2}-1}}{\sqrt{1+\Lambda ^{2}}}\exp %
\left[ -\frac{r_{\bot }^{2}}{w_{0}^{2}\left( 1+\Lambda ^{2}\right) }\right]
\end{equation*}%
\begin{equation}
\times f\left( x^{\prime }\right) \cos \left( k^{\prime }x^{\prime }-\tan
^{-1}\left( \Lambda \right) \right) ,  \label{A42}
\end{equation}%
where%
\begin{equation*}
\Lambda =\frac{\left( x^{\prime }+\frac{c}{n_{0}}t^{\prime }\right) }{%
x_{R}^{\prime }}.
\end{equation*}%
Here $x_{R}^{\prime }=x_{R}/\gamma $ is the Lorentz contracted Rayleigh
range and $k^{\prime }=\omega \sqrt{n_{0}^{2}-1}/c$. In this frame the
characteristic motion of electrons is nonrelativistic. If the Cherenkov
angle is larger than the laser beam diffraction angle $\vartheta
_{0}=w_{0}/x_{R}$, or interaction time is smaller than $t_{R}=n_{0}\pi
w_{0}^{2}\sqrt{n_{0}^{2}-1}/\lambda c$ one will have 
\begin{equation}
H_{y}^{\prime }=-E_{0}\sqrt{n_{0}^{2}-1}\exp \left[ -\frac{r_{\bot }^{2}}{%
w_{0}^{2}}\right] f\left( x^{\prime }\right) \cos \left( k^{\prime
}x^{\prime }\right) ,  \label{A6}
\end{equation}

\section{Classical Analysis of the Induced Cherenkov Process: Smooth turn on
and off the interaction}

We have numerically solved the classical equations of motion in the $R$
frame for an ensemble of electrons subjected to the magnetic field of a
slowed traveling wave: 
\begin{equation*}
\frac{d\mathrm{v}_{x}}{dt}=-\frac{e\mathrm{v}_{z}H_{y}}{mc};\ \frac{dx}{dt}=%
\mathrm{v}_{x},
\end{equation*}%
\begin{equation}
\frac{d\mathrm{v}_{z}}{dt}=\frac{e\mathrm{v}_{x}H_{y}}{mc};\frac{dz}{dt}=%
\mathrm{v}_{z}.  \label{C1}
\end{equation}%
The wave magnetic field is assumed to have a Gaussian transverse profile
(the Cherenkov angle is larger than the laser beam diffraction angle): 
\begin{equation*}
\mathbf{H}(x,z)=-\hat{y}E_{0}\sqrt{n_{0}^{2}-1}\,e^{-z^{2}/w_{0}^{2}}f(x)%
\cos (k^{\prime }x),
\end{equation*}%
with a beam waist of $w_{0}=100\lambda $.

We consider an ensemble of $10^{3}$ electrons uniformly distributed along a
phase lattice with $N_{k}=10$ periods. The initial transverse coordinate is
fixed at $z_{0}=-250\lambda $, and the initial longitudinal velocity is set
to $\mathrm{v}_{0x}=0$, consistent with the classical Cherenkov resonance
condition. The transverse momentum $p_{0z}=0.1\,mc$ is Lorentz invariant and
matches the quantum case, corresponding to an initial velocity $\mathrm{v}%
_{0z}/c=0.1$.

\begin{figure}[tbp]
\includegraphics[width=0.48\textwidth]{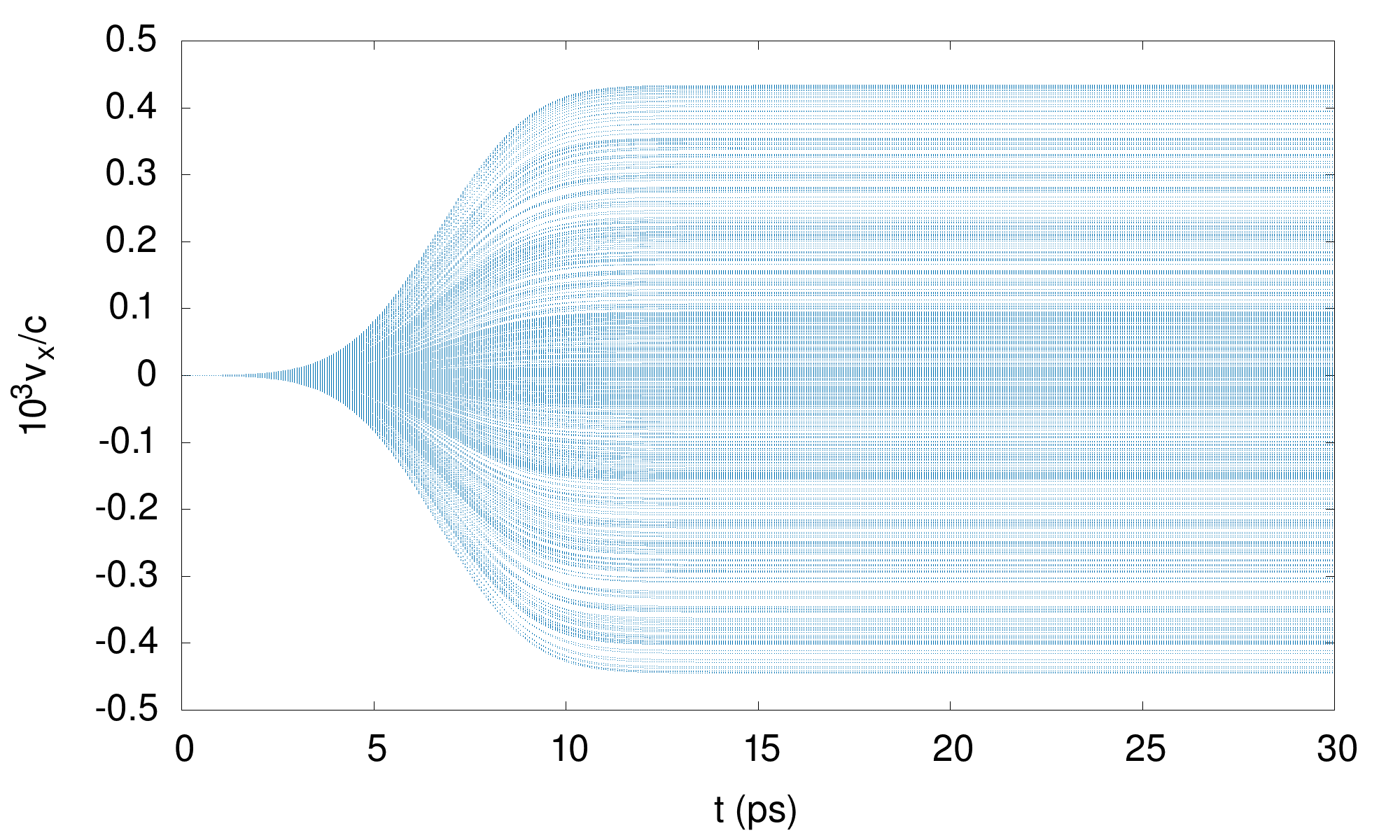}
\caption{Classical analysis of the induced Cherenkov process. Longitudinal
velocity distribution of an electron ensemble at $\protect\xi %
_{0}=1.25\times 10^{-5}$, as obtained from Eq. (\protect\ref{C1}).}
\label{figC1}
\end{figure}

\begin{figure}[tbp]
\includegraphics[width=0.48\textwidth]{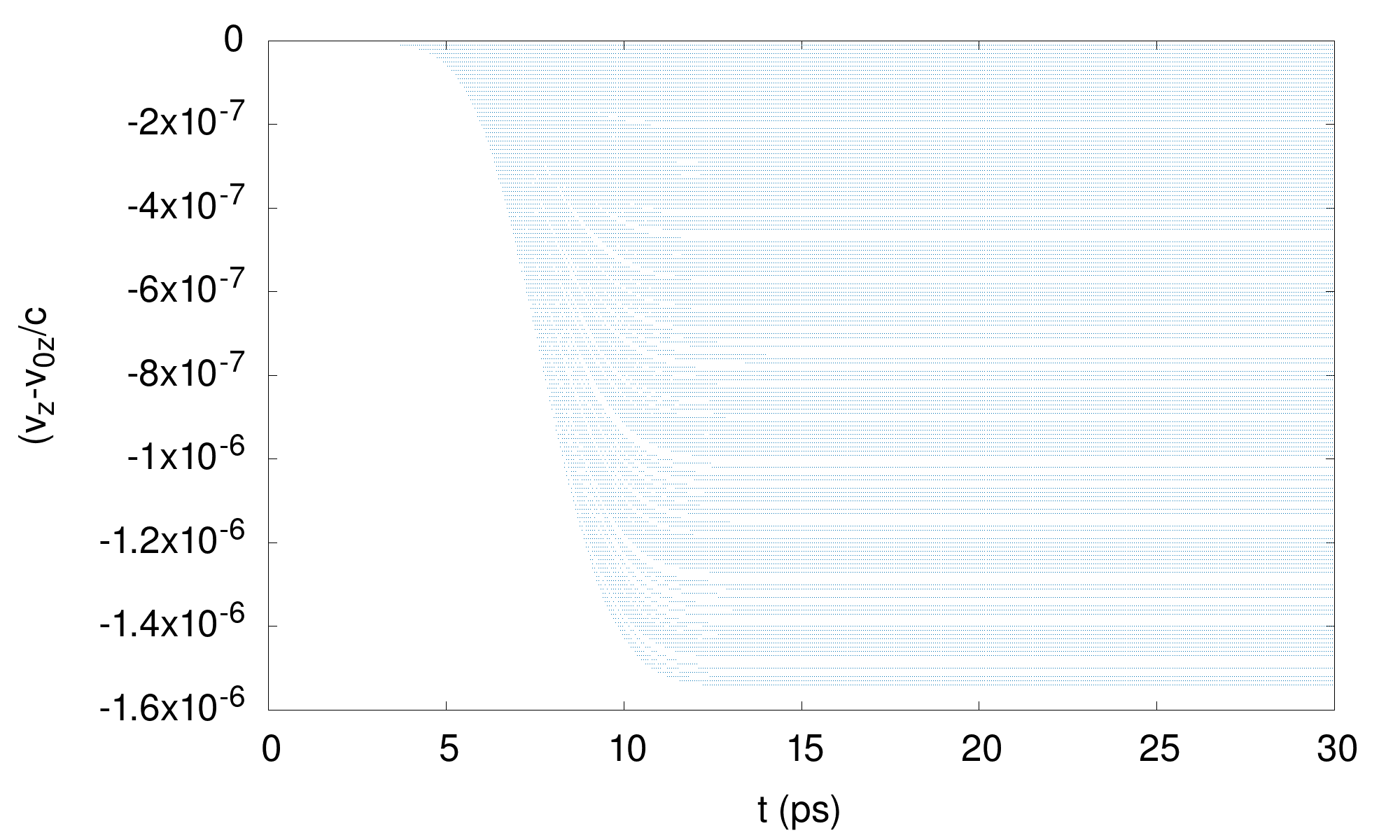}
\caption{Transverse velocity change according to Eq. (\protect\ref{C1}). The
change is negligible, confirming the validity of a free-particle
approximation in the transverse direction.}
\label{figC2}
\end{figure}

Figure \ref{figC1} shows the distribution of the final longitudinal
velocities across the ensemble. As expected, the electrons acquire
velocities ranging from $-\mathrm{v}_{x\max }$ to $\mathrm{v}_{x\max }$,
depending on their initial position. Figure \ref{figC2} shows the change in
transverse velocity, which remains negligible for all electrons. Hence, the
transverse motion is effectively free and follows 
\begin{equation*}
z(t)=z_{0}+\mathrm{v}_{0z}t.
\end{equation*}

This implies that the field experienced by the electrons can be modeled as $%
\mathbf{H}(x,z)\rightarrow \mathbf{H}(x,z_{0}+\mathrm{v}_{0z}t)$,
effectively introducing a smooth temporal envelope. For a laser wavelength
of $\lambda =800\,\mathrm{nm}$, this yields a Gaussian envelope function 
\begin{equation*}
g(t)=\exp \left[ -\left( t-2.5\tau _{w}\right) ^{2}/\tau _{w}^{2}\right] ,
\end{equation*}%
with $\tau _{w}=w_{0}/\mathrm{v}_{0z}=2.67\,\mathrm{ps}$. This provides
physical justification for the use of a Gaussian envelope in the quantum
analysis, which is valid as long as the electron's transverse position
uncertainty satisfies $\delta z\ll 2w_{0}$. Assuming $\delta z\sim \delta x$%
, this condition is well met.

For weak laser fields, an analytical expression for the longitudinal
velocity can be obtained from Eq. (\ref{C1}):%
\begin{equation*}
\mathrm{v}_{x}(t)=\frac{ev_{0z}k^{\prime }A_{0}}{mc}
\end{equation*}%
\begin{equation}
\times f(x_{0})\cos (k^{\prime
}x_{0})\int_{0}^{t}e^{-(z_{0}+v_{0z}t)^{2}/w_{0}^{2}}dt.  \label{Cv}
\end{equation}%
For $t>2.5\tau _{w}$, this saturates to: 
\begin{equation*}
\mathrm{v}_{x}(x_{0})=\mathrm{v}_{x\max }f(x_{0})\cos (k^{\prime }x_{0}),
\end{equation*}%
where the maximal velocity amplitude is: 
\begin{equation}
\mathrm{v}_{x\max }=\sqrt{\pi }ck^{\prime }\xi _{0}w_{0}.  \label{C2}
\end{equation}

This result establishes a correspondence between the classical and quantum
descriptions in the regime of large photon absorption/emission. In the
quantum picture, the most probable number of absorbed/emitted Cherenkov
photons at $Z_{B}\gg 1$ is $s=Z_{B}$, leading to a most probable velocity: 
\begin{equation}
\mathrm{v}_{x,\mathrm{prob}}=\frac{\hbar k^{\prime }Z_{B}}{m}=\mathrm{v}%
_{x\max }.  \label{C3}
\end{equation}



\bibliographystyle{model3-num-names.bst}
\bibliography{bibliography.bib}






\end{document}